\documentclass[fleqn,10pt]{wlscirep}
\usepackage[utf8]{inputenc}
\usepackage[T1]{fontenc}
\usepackage{xcolor,colortbl}
\usepackage[numbers]{natbib}
\usepackage{multirow,enumitem,enumerate}
\usepackage{amssymb} 
\usepackage{tikz}
\usepackage{graphicx}
\usepackage{microtype}
\usepackage{booktabs} 
\usepackage{multirow}
\usepackage{amssymb}
\usepackage{mathtools}
\usepackage{amsthm}
\usepackage{arydshln}
\usepackage{pifont}
\usepackage[capitalize,noabbrev]{cleveref}
\usepackage{url}

\usepackage{breakurl}

\usepackage[textsize=tiny]{todonotes}

\definecolor{markChanges}{rgb}{1,0,1}

\usepackage{lineno}

\usepackage{lipsum}

\newcommand\blfootnote[1]{%
  \begingroup
  \renewcommand\thefootnote{}\footnote{#1}%
  \addtocounter{footnote}{-1}%
  \endgroup
}

\title{Coswara: A respiratory sounds and symptoms dataset for remote screening of SARS-CoV-2 infection}

\author[1]{Debarpan Bhattacharya}
\author[2]{Neeraj Kumar Sharma}
\author[1]{Debottam Dutta}
\author[3]{Srikanth Raj Chetupalli}
\author[1]{Pravin Mote}
\author[1,*]{Sriram Ganapathy}
\author[4]{Chandrakiran C} 
\author[4]{Sahiti Nori} 
\author[4]{Suhail K K} 
\author[5]{Sadhana Gonuguntla} 
\author[6]{Murali Alagesan}   

\affil[1]{Department of Electrical Engineering, Indian Institute of Science, Bangalore, India} 
\affil[2]{Mehta Family School of Data Science and Artificial Intelligence, Indian Institute of Technology, Guwahati, India} 
\affil[3]{Fraunhofer Institute of Integrated Circuits, Erlangen, Germany}
\affil[4]{Ramaiah Medical College Hospital, Bangalore, India}
\affil[5]{General Hospital, Hoskote, Bangalore, India}
\affil[6]{PSG Institute of Medical Sciences and Research, Coimbatore, India} 

\affil[*]{sriramg@iisc.ac.in}

\begin{abstract}
This paper presents the Coswara dataset, a dataset containing diverse set of respiratory sounds and rich meta-data, recorded between April-2020 and February-2022 from 2635 individuals (1819 SARS-CoV-2 negative, 674 positive, and 142 recovered subjects). The respiratory sounds contained nine sound categories associated with variants of breathing, cough and speech. The rich metadata contained demographic information associated with age, gender and geographic location, as well as the health  information relating to the symptoms, pre-existing respiratory ailments, comorbidity and SARS-CoV-2 test status. Our study is the first of its kind to manually annotate the audio quality of the entire dataset (amounting to 65~hours) through manual listening. The paper summarizes the data collection procedure, demographic, symptoms and audio data information. A COVID-19 classifier based on bi-directional long short-term (BLSTM) architecture, is trained and evaluated on the different population sub-groups contained in the dataset to understand the bias/fairness of the model. This enabled the analysis of the impact of gender, geographic location, date of recording, and language proficiency on the COVID-19 detection performance.
\end{abstract}
\begin{document}

\flushbottom
\maketitle

\thispagestyle{empty}

\blfootnote{\noindent\textbf{Accepted for publication in Nature Scientific Data, 2023}}

\section*{Background \& Summary}
As of July 2022, with more than $550$ million reported COVID-19 cases and a fatality ratio of more than $1\%$, the COVID-19 pandemic has emerged as the most consequential global health crisis since the influenza pandemic of $1918$ \cite{cascella2022features}. The outbreak has largely outpaced global efforts to characterize the infection and contain its spread. 
The measures such as population surveillance, case identification, contact tracing, vaccination, physical distancing, mask mandates, and lock-downs have helped control the outbreak to a certain extent.
Inventing alternative COVID-19 screening methodologies, which can function as point-of-care tests (POCT), and are efficient in terms of time, cost and performance was highlighted as an urgent requirement by the World Health Organization (WHO)
[\href{https://www.who.int/docs/default-source/blue-print/who-rd-blueprint-diagnostics-tpp-final-v1-0-28-09-jc-ppc-final-cmp92616a80172344e4be0edf315b582021.pdf?sfvrsn=e3747f20_1&download=true}{link}]
The application of digital technologies for large-scale health-related data collection and analysis can help meet this requirement \cite{budd2020digital, lipsitch2020defining} and build infrastructure to tackle future pandemics. The longitudinal studies, recording self-reported symptoms using smartphone-based applications, had explored the development of easily accessible COVID-19 screening tools \cite{drew2020rapid, menni2020real, zoabi2021machine, natarajan2020assessment}. Besides symptoms, measurement of physiological signals such as respiration rate using wearable electronic sensors had also been explored for COVID-19 detection \cite{natarajan2021measurement}.

The COVID-19 is primarily a respiratory illness \cite{hu2020characteristics}. Listening to respiratory sounds, such as deep breathing, with a stethoscope has served as a useful methodology to screen  respiratory ailments since 1821 \cite{laennec1838treatise}.
Currently, utilizing digital technologies, respiratory sound samples can be collected via internet connected devices. The computer aided analysis of respiratory sounds can bring new insights for the 
design of POCT solutions as well as in the monitoring of SARS-CoV-2 infection.
Motivated by this, we conceptualized the creation of a respiratory sound and symptom dataset named 
\textit{Coswara} (an amalgamation of ``\text{Co}'' from COVID-19 and ``\text{swara}'' meaning sound in Sanskrit),
composed of breathing, cough, and speech sounds along with the  health related symptoms, recorded from individuals with, without, and recovered from SARS-CoV-2 infection \cite{sharma2020coswara}. This paper presents a detailed description of the data collection protocol, a summary of the data records, and results from bias analysis performed with a pre-trained classifier model.

In comparison with other efforts on respiratory sound sample collection  \cite{orlandic2020coughvid, han2022sounds, pizzo2021iatos}, the Coswara dataset was primarily collected from India. Further, the data records were collected between April-2020 to February-2022, allowing the Coswara dataset to contain records associated with infections due to multiple variants of the SARS-CoV-2 virus. The participant metadata provided in the Coswara dataset spanned a wide range of attributes and included demographic, health, symptom, and COVID-19 status information. For respiratory sound recordings, each participant provided nine sound recordings. These correspond to   breathing (breathing deep and breathing shallow), cough (coughing deep and coughing shallow), sustained vowel phonation (three different vowels), and continuous speech (counting from $1$- $20$ at normal and fast pace). These sound recordings were released without application of lossy compression standards. In order to validate the audio files that were collected using the crowd-sourced web-link, the human annotators listened to all the sound recordings and marked the quality in a scale of $1$-$3$. The sound quality annotations obtained from human listening were also released as part of the dataset. A detailed comparison of the Coswara dataset with few other relevant dataset is provided in Table~\ref{tab:other_datasets}. As shown in this Table, the Coswara dataset is unique in terms of containing nine different types of sound samples, augmented with a rich meta-data, audio quality annotation obtained through human listening. The dataset also contains labels associated with COVID-19 recovered subject population.
A COVID-19 screening tool designed using this dataset is made publicly available. Furthermore, we report a detailed bias and fairness analysis to understand the confounding factors associated with the COVID-19 screening tool developed using the collected data.


\begin{figure*}[t!]
    \centering
    \includegraphics[width=\textwidth]{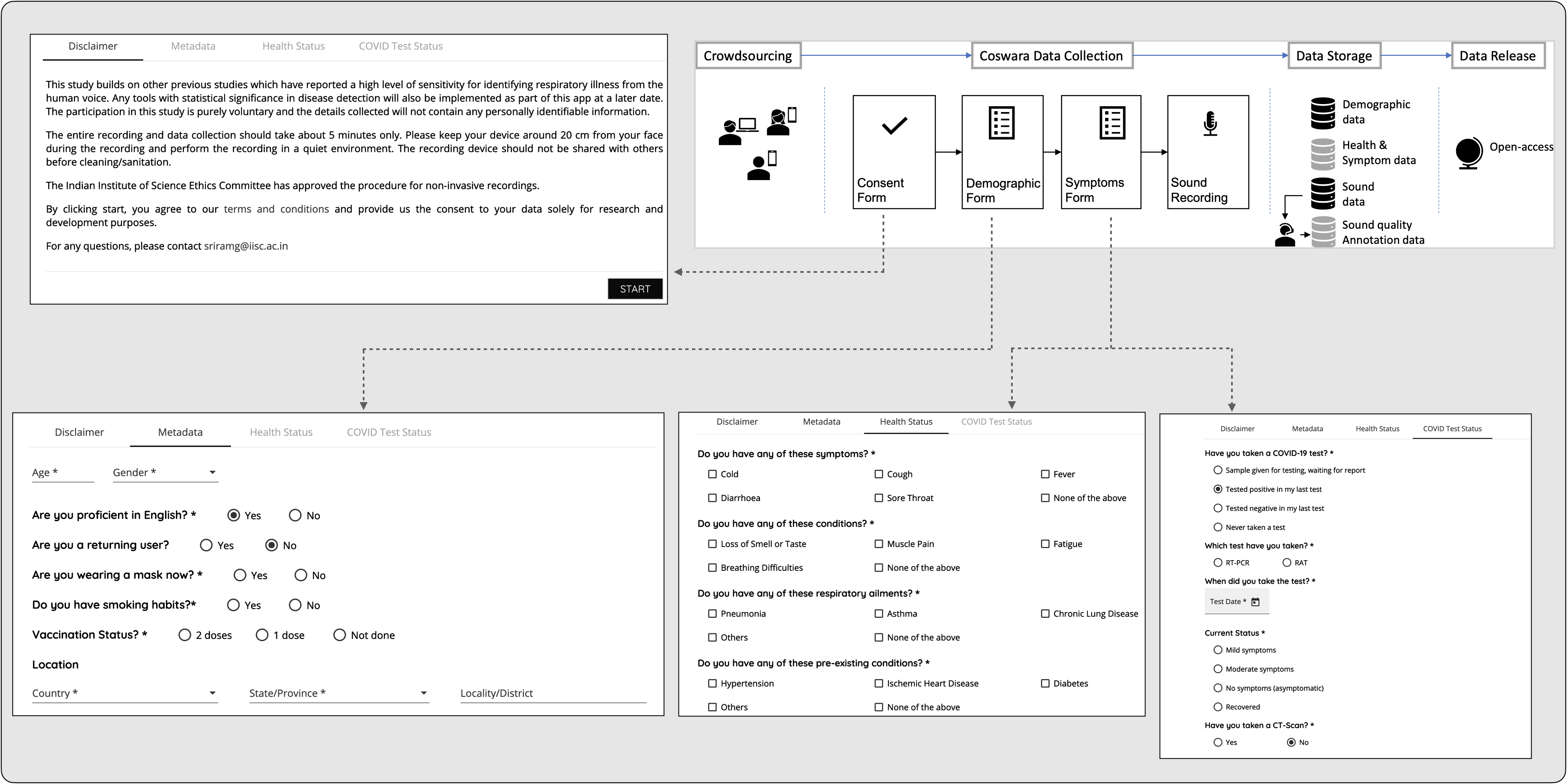}
    \caption{A block diagram illustration of the steps involved in creating the Coswara dataset.}
    \vspace{-0.1in}
    \label{fig:dataset_creation}
\end{figure*}
\begin{table*}[!t]
\resizebox{\textwidth}{!}
{
\begin{tabular}{@{}llllllllllclclllrlrlrlc@{}}
\toprule
\multirow{2}{*}{\textbf{\begin{tabular}[c]{@{}l@{}}Dataset\\ Name\end{tabular}}} & \multirow{2}{*}{\textbf{}} & \multirow{2}{*}{\textbf{\begin{tabular}[c]{@{}l@{}}Data\\ Collected\\ From\end{tabular}}} & \multirow{2}{*}{\textbf{}} & \multirow{2}{*}{\textbf{Data Access}} & \multirow{2}{*}{\textbf{}} & \multirow{2}{*}{\textbf{Meta-data}} & \multirow{2}{*}{\textbf{}} & \multirow{2}{*}{\textbf{Sound Data}} & \multirow{2}{*}{\textbf{}} & \multicolumn{1}{l}{\multirow{2}{*}{\textbf{\begin{tabular}[c]{@{}l@{}}Sound\\ Samples\\ per Subject\end{tabular}}}} & \multirow{2}{*}{\textbf{}} & \multicolumn{1}{l}{\multirow{2}{*}{\textbf{\begin{tabular}[c]{@{}l@{}}Uncompressed\\ audio files\end{tabular}}}} & \multirow{2}{*}{\textbf{}} & \multirow{2}{*}{\textbf{\begin{tabular}[c]{@{}l@{}}Audio\\ Quality\\ Validation\end{tabular}}} & \multirow{2}{*}{\textbf{}} & \multicolumn{5}{c}{\textbf{Subject Count}} &  & \multicolumn{1}{l}{\multirow{2}{*}{\textbf{\begin{tabular}[c]{@{}l@{}}COVID-19\\ Screening\\ Tool\\ Released\end{tabular}}}} \\ \cmidrule(lr){17-22}
 &  &  &  &  &  &  &  &  &  & \multicolumn{1}{l}{} &  & \multicolumn{1}{l}{} &  &  &  & \textbf{Non-COVID-19} & \multicolumn{1}{r}{\textbf{}} & \textbf{COVID-19} & \multicolumn{1}{r}{\textbf{}} & \textbf{\begin{tabular}[c]{@{}r@{}}Recovered\\ from\\ COVID-19\end{tabular}} &  & \multicolumn{1}{l}{} \\ \cmidrule(r){1-1} \cmidrule(lr){3-3} \cmidrule(lr){5-5} \cmidrule(lr){7-7} \cmidrule(lr){9-9} \cmidrule(lr){11-11} \cmidrule(lr){13-13} \cmidrule(lr){15-15} \cmidrule(lr){17-17} \cmidrule(lr){19-19} \cmidrule(lr){21-21} \cmidrule(l){23-23} 
\begin{tabular}[c]{@{}l@{}}COUGHVID\\\cite{orlandic2021coughvid}\end{tabular} &  & \begin{tabular}[c]{@{}l@{}}EU and\\ beyond\end{tabular} &  & Open &  & \begin{tabular}[c]{@{}l@{}}Age\\ Gender\\ Country\\ Symptoms\\ Medical history\end{tabular} &  & Cough &  & 1 &  & $\times$ &  & $\times$ &  & 26395 &  & 1155 &  & $\times$ &  & $\times$ \\ \cmidrule(r){1-1} \cmidrule(l){3-23} 
\begin{tabular}[c]{@{}l@{}}COVID-19\\ Sounds\\\cite{xia2021covid}\end{tabular} &  & \begin{tabular}[c]{@{}l@{}}EU, USA,\\ and beyond\end{tabular} &  & On request &  & \begin{tabular}[c]{@{}l@{}}Gender\\ Symptoms\\ Medical history\end{tabular} &  & \begin{tabular}[c]{@{}l@{}}Cough\\ Breathing\\ Speech\end{tabular} &  & 3 &  & $\times$ &  & Automatic &  & 6450 &  & 2106 &  & $\times$ &  & $\times$ \\ \cmidrule(r){1-1} \cmidrule(l){3-23} 
\begin{tabular}[c]{@{}l@{}}Tos\\ COVID-19\\\cite{pizzo2021iatos}\end{tabular} &  & Argentina &  & Open &  & \begin{tabular}[c]{@{}l@{}}Age\\ Gender\\ Location\\ Symptoms\end{tabular} &  & Cough &  & 1 &  & $\times$ &  & $\times$ &  & 125183 &  & 21197 &  & $\times$ &  & $\times$ \\ \cmidrule(r){1-1} \cmidrule(l){3-23} 
\begin{tabular}[c]{@{}l@{}}Coswara\\ (This work)\end{tabular} &  & \begin{tabular}[c]{@{}l@{}}India and\\ beyond\end{tabular} &  & Open &  & \begin{tabular}[c]{@{}l@{}}Age\\ Gender\\ State\\ Country\\ Symptoms\\ Medical history\\ (includes\\ respiratory health,\\ COVID-19\\ vaccination status,\\ other diseases,\\ and more)\end{tabular} &  & \begin{tabular}[c]{@{}l@{}}Cough\\ (2 types)\\ Breathing\\ (2 types)\\ Vowel\\ (3 types)\\ Speech\\ (2 types)\end{tabular} &  & 9 &  & $\checkmark$ &  & \begin{tabular}[c]{@{}l@{}}Manual\\ Listening\end{tabular} &  & 1819 &  & 674 &  & 142 &  & $\checkmark$ \\ \bottomrule
\end{tabular}
}
\caption{List of publicly accessible COVID-19 related respiratory sound datasets.}
\label{tab:other_datasets}
\vspace{-0.05in}
\end{table*}


India shares about 18\% of the global population and features approximately 32\% of  the global disability adjusted life years (DALY, an estimate of overall disease burden) from respiratory diseases \cite{salvi2018burden}. In addition to the analysis of the health symptoms and respiratory acoustics associated with COVID-19 disease, we foresee  the Coswara dataset to be of future relevance to understand respiratory acoustics and aid in the design of cost-effective, scalable, and accurate respiratory health monitoring technologies.



\section*{Methods}
\subsection*{Design}
The Coswara dataset was based on crowdsourcing via a website
(\url{https://coswara.iisc.ac.in/}).
The application interface was designed with a simple workflow (see Figure~\ref{fig:dataset_creation}). Any individual connected to the internet could access the website using a smartphone (or a computer). The first step was the collection of the  participant's  consent to record their audio. Subsequently, the participant filled out a short questionnaire to provide demographic information, current health status, and COVID-19 test results.
A detailed list of all the metadata recorded is provided in Table~\ref{tab:collected_datasheet}. The participant  recorded nine different kinds of sound samples. A description of the sound samples recorded is provided in Table~\ref{tab:sound_categories}.

\subsubsection*{Demographics data}
The participants provided demographic information like  age, gender, country and the   location information in terms of province. Further, they also indicated whether they were  proficient in English, and whether they had a face mask during the recording. The smoking status was also recorded. 

\subsubsection*{Symptom data}
The participants provided information about the presence of COVID-19-like symptoms, respiratory ailments, comorbidity, and COVID-19 test status. For each of these, a list of options was provided and the participants chose as many options as were applicable to them. A summary of the list of  symptoms provided by the participants is shown in Table~\ref{tab:collected_datasheet}.

\subsubsection*{COVID-19 test status}
The participants provided information on their COVID-19 status by classifying themselves into one of the three categories, namely, COVID-19 negative (non-COVID), positive (COVID), and recovered categories. If the participant belonged to a positive or recovered category, they further provided the date of the last COVID-19 test conducted and the kind of test (RAT or RT-PCR).

\subsubsection*{Respiratory sound data}
The website application provided a sound recording interface. The participants were instructed to record and upload nine sound samples, sequentially as different audio files. An example sound sample for each sound category was also provided. The nine sound samples included two variants of breathing, two variants of cough, three variants of sustained vowel phonation, and two variants of continuous speech. A description of these sound categories is provided in Table~\ref{tab:sound_categories}. 
\begin{table}[t!]
\centering
\begin{tabular}{cllll}
\hline
\multicolumn{1}{l}{\textbf{\begin{tabular}[c]{@{}l@{}}Sound\\ Category\end{tabular}}} &  & \textbf{\begin{tabular}[c]{@{}l@{}}Collected\\ Sound Sample\end{tabular}} & \multicolumn{1}{c}{} & \textbf{Description} \\ \cline{1-1} \cline{3-3} \cline{5-5} 
\multicolumn{1}{l}{} &  &  &  &  \\
\multirow{2}{*}{Breathing} &  & Breathing-shallow &  & \begin{tabular}[c]{@{}l@{}}Few cycles of inspiration and expiration without exertion on the lungs\end{tabular} \\
 &  & Breathing-deep &  & \begin{tabular}[c]{@{}l@{}}Few cycles of inspiration and expiration with exertion on the lungs\end{tabular} \\ \cline{1-1} \cline{3-3} \cline{5-5} 
\multicolumn{1}{l}{} &  &  &  &  \\
\multirow{2}{*}{Cough} &  & Cough-shallow &  & \begin{tabular}[c]{@{}l@{}}Few bouts of (induced) cough without exertion on lungs\end{tabular} \\
 &  & Cough-heavy &  & \begin{tabular}[c]{@{}l@{}}Few bouts of (induced) cough with exertion on lungs\end{tabular} \\ \cline{1-1} \cline{3-3} \cline{5-5} 
\multicolumn{1}{l}{} &  &  &  &  \\
\multirow{3}{*}{\begin{tabular}[c]{@{}c@{}}Vowel\\ Phonation\end{tabular}} &  & Vowel-[u] &  & \begin{tabular}[c]{@{}l@{}}Sustained phonation of vowel-[u] (as in the word boot)\end{tabular} \\
 &  & Vowel-[i] &  & \begin{tabular}[c]{@{}l@{}}Sustained phonation of vowel-[i] (as in the word beet)\end{tabular} \\
 &  & Vowel-[\ae] &  & \begin{tabular}[c]{@{}l@{}}Sustained phonation of vowel-[\ae] (as in the word bat)\end{tabular} \\ \cline{1-1} \cline{3-3} \cline{5-5} 
\multicolumn{1}{l}{} &  &  &  &  \\
\multirow{2}{*}{Speech} &  & Counting-normal &  & \begin{tabular}[c]{@{}l@{}}Counting numbers from 1 to 20 in normal pace\end{tabular} \\
 &  & Counting-fast &  & \begin{tabular}[c]{@{}l@{}}Counting numbers from 1 to 20 in fast pace\end{tabular} \\ \hline
\end{tabular}

\caption{Description of the sound categories in the Coswara dataset.}
\vspace{-0.05in}
\label{tab:sound_categories}
\end{table}
\subsection*{Procedures}
The website URL was shared with the public through various social media platforms and word-of-mouth in the collaborating hospitals and health centers. The crowd-sourcing approach to data collection enabled the recording of data from diverse age groups, geographic locations (within India and to a smaller extent, from outside India), health conditions, sound recording device types and ambient environments. A participant's complete data record comprised of demographic, symptom, COVID status, and sound data.   On average, each participant spent approximately $7$~minutes to record their data.
A significant portion ($> 95$\%) of the subjects who were SARS-CoV-2 infected, were inducted through hospitals and health centers. 
Hence, the label quality of these data records were validated. 

\subsection*{Ethical issues}
The data collection procedure was approved by the Institutional Human Ethics Committee, at the Indian Institute of Science, Bangalore. The informed consent was obtained from all participants who uploaded their data records. All the data collected was anonymized, and excluded any participant identity information. 

\section*{Data records}
\subsection*{Manual curation}
\noindent Approximately $65$ hours of respiratory sound samples ($23700$ recordings coming from 2635 subjects each having 9 categories of recordings) were subjected to manual listening by the human annotators. The listeners classified the recordings based on the sound quality into one of the three categories, namely, excellent (no ambient noise), moderate (slight ambient noise), and poor (significant ambient noise and distortions). The curation process resulted in identifying $78.6\%$ of the samples as excellent, $11.7\%$ as moderate, and $9.7\%$ as poor quality sound recordings, respectively.\\
The Coswara dataset is publicly available as a Zenodo repository~\cite{debarpan_bhattacharya_2022_7188627}.

\begin{figure}[t!]
    \centering
    \input{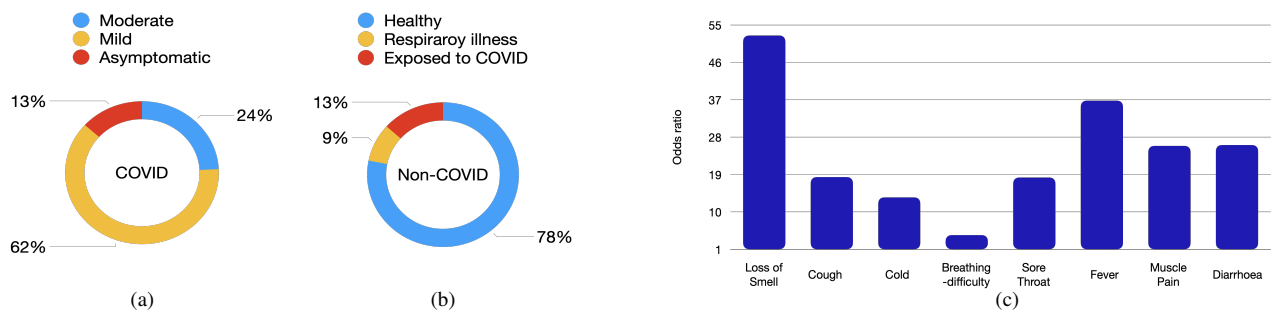}
   
    \caption{Distribution of the (a) COVID, and (b) non-COVID subjects corresponding to different sub-categories. Figure (c) shows the odds ratio for the different symptoms. }
    \label{fig:plot_subject_covid_stat}

\end{figure}

\section*{Technical validation}
\subsection*{COVID-19 representation}
From the perspective of SARS-CoV-2 infection, the dataset was  comprised of three kinds of subject categories (see Figure~\ref{fig:plot_subject_covid_stat}(a)). First, the \textit{non-COVID} category contained data records from $1809$ subjects. These subjects were completely healthy ($78\%$), had some respiratory ailments ($9\%$), or had COVID-19 like symptoms ($13\%$). Further, a subset of these participants self-reported themselves as exposed to the virus ($13\%$), but  had not tested positive for SARS-CoV-2 infection in the past or at the time of data recording. Second, the \textit{COVID} category contained data records from $674$ subjects. These individuals had tested positive for SARS-CoV-2 infection at the time of data recording or in the past $14$ days. Using the self-reported health condition, these individuals were further grouped into asymptomatic ($14\%$), mild ($62\%$) and moderate ($24\%$) COVID-19 patients. Third, the \textit{recovered} category contained data records from $142$ subjects. These subjects had completed at least $14$ days since testing positive for COVID-19.

\subsection*{Demographic representation}
\begin{figure}[t!]
    \centering
    \input{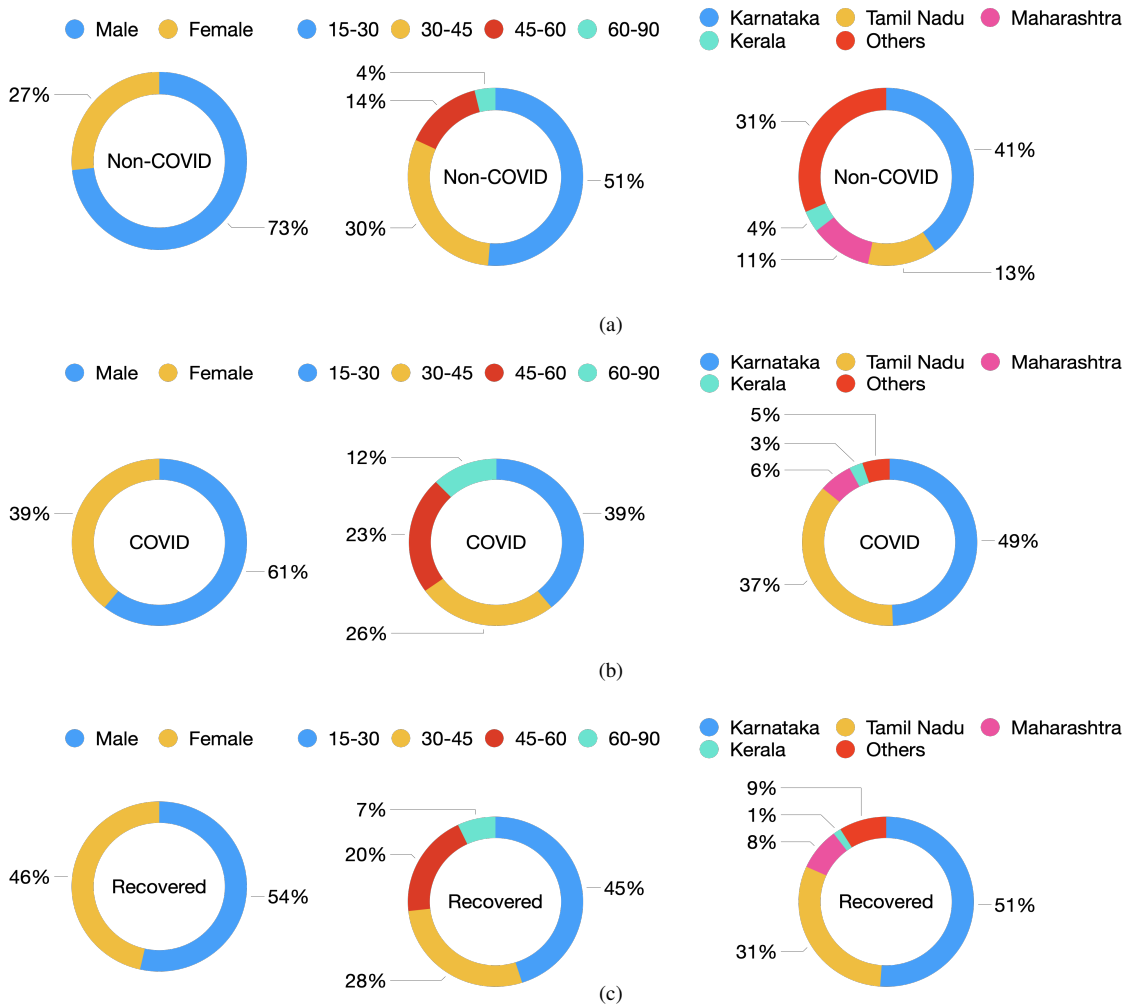}
    \vspace{-0.05in}
    \caption{Distribution of the (a) non-COVID (1819 subjects), (b) COVID-19 positive (674 subjects), and (c) recovered from COVID-19 (142 subjects) population in terms of gender, age and location.}
    \label{fig:plot_metadata_pie}
\end{figure}

\noindent The demographic distribution for each subject category, namely, non-COVID, COVID and recovered, is shown in Figure~\ref{fig:plot_metadata_pie}.
The participating subjects spanned a wide age group (between $15-90$ years), with a majority between $15-45$ years. Further, the dataset also contained more male participants. The geographic distribution of the data was concentrated primarily in India ($91\%$ from India). Within India, a majority of the data came from Karnataka, a province in southern India. The rest of the data was drawn from various other provinces across India.

\subsection*{Symptom representation}
\begin{figure}[!h]
    \centering
    \input{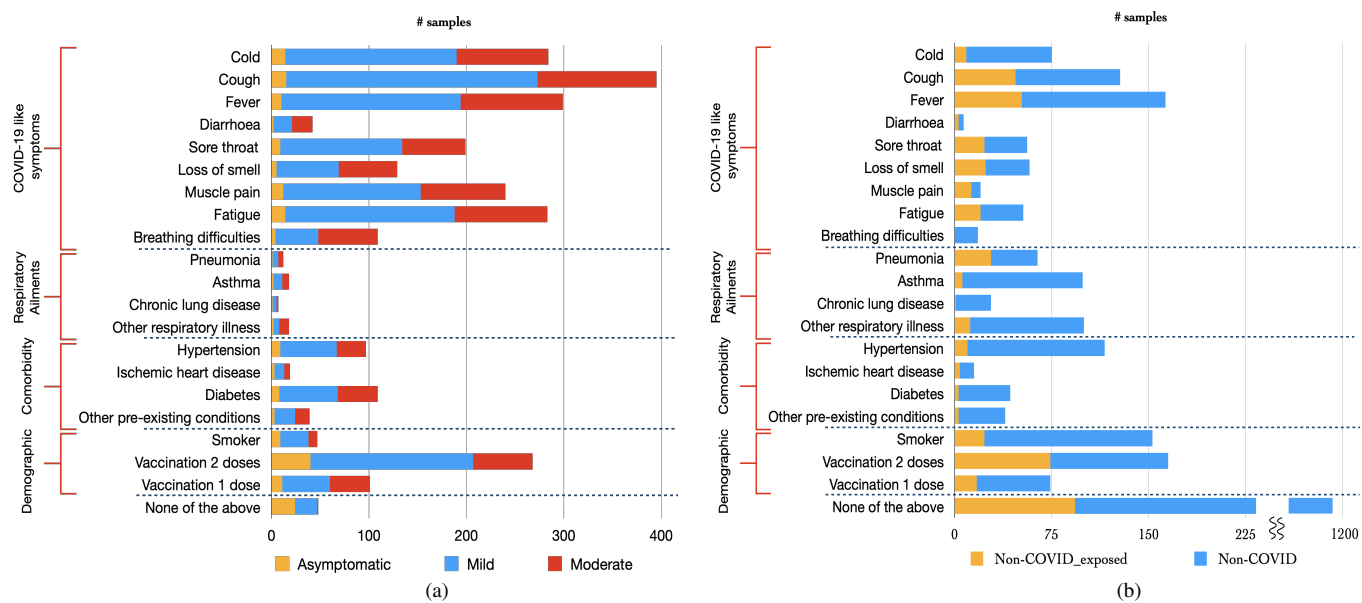}
    \vspace{-0.05in}
    \caption{(a) COVID-19 positive and (b) non-COVID subject meta data distribution.}
    \label{fig:plot_metadata_pon_neg}
\vspace{-0.1in}
\end{figure}

The distribution of the symptoms among the non-COVID and COVID subjects is shown in Figure~\ref{fig:plot_metadata_pon_neg}. As expected, the COVID-like symptoms such as cold, cough, fever and fatigue, were relatively more prevalent in COVID subjects compared to non-COVID. In the non-COVID class, a majority of the subjects were healthy, i.e., without any COVID-19-like symptoms, respiratory ailments, or comorbidity. However, there were considerable number of non-COVID subjects with COVID-19-like symptoms of fever ($225$ individuals), cough ($173$ individuals) and cold ($90$ individuals). Further, there were also several non-COVID subjects with pre-existing respiratory ailments of pneumonia ($83$ individuals), Asthma ($110$ individuals) and other respiratory illnesses ($120$ individuals). Hence, the non-COVID category data broadly represents the population level incidence of different respiratory ailments. Some of the subjects, also indicated a vaccination status of $2$ doses. 

Figure~\ref{fig:plot_subject_covid_stat}(c) depicts the odds ratio, a statistic that quantifies the strength of the association between health status and a symptom. It is defined as the ratio of the odds of ``symptom $x$'' in the COVID category to the odds of ``symptom $x$'' in the non-COVID category. Here, we see that most of the COVID-19 like-symptoms had an odds ratio $>1$.

\begin{table}[]
\resizebox{\columnwidth}{!}
{
\begin{tabular}{@{}llll@{}}
\toprule
\textbf{Metadata Type} & \textbf{Field name} & \textbf{Description} & \textbf{Allowed values} \\ \midrule
\multirow{11}{*}{Demographic} & id & participant identifier & assigned using a unique random code generator \\
 & record\_data & data recording date & dd-mm-yy \\
 & a & age & number \\
 & g & gender & male/female/other \\
 & l\_c & country & one from a list of 217 countries \\
 & l\_s & state & one from a list of province associated with l\_c \\
 & vacc & COVID-19 vaccination status & y (at least two doses)/p (one dose)/n (unvaccinated) \\
 & ep & proficient in english & True/False \\
 & smoker & regular smoker & True/False \\
 & rU & returning participant & True/False \\
 & um & wearing mask & True/False \\
 &  &  &  \\ \midrule
\multirow{9}{*}{\begin{tabular}[c]{@{}l@{}}COVID-19-like\\ symptoms\end{tabular}} & cough & has cough & True/False \\
 & cold & has cold & True/False \\
 & diarrhoea & has diarrhoea & True/False \\
 & bd & has breathing difficulties & True/False \\
 & st & has sore throat & True/False \\
 & fever & has fever & True/False \\
 & ftg & suffering from fatigue & True/False \\
 & mp & has muscle pain & True/False \\
 & has loss\_of\_smell & loss of smell and/or taste & True/False \\
 &  &  &  \\ \midrule
\multirow{4}{*}{\begin{tabular}[c]{@{}l@{}}Respiratory\\ ailments\end{tabular}} & asthma & has asthma related issues & yes/no \\
 & cld & has chronic lung disease & True/False \\
 & pneumonia & has pneumonia & True/False \\
 & others\_resp & has other respiratory illness & True/False \\
 &  &  &  \\ \midrule
\multirow{4}{*}{Comorbidity} & ht & has hypertension & True/False \\
 & diabetes & has diabetes & True/False \\
 & ihd & has ischemic heart disease & True/False \\
 & others\_preexist & any other pre-existing comorbidity & True/False \\
 &  &  &  \\ \midrule
\multirow{7}{*}{COVID-19 health} & test\_status & status of COVID-19 test & p (positive)/n (negative)/na (not taken a test) \\
 & covid\_status & covid related health status & \begin{tabular}[c]{@{}l@{}}positive\_mild, healthy, positive\_moderate,\\ positive\_asymptomatic, exposed\end{tabular} \\
 & testType & type of COVID-19 test taken & RAT/RT-PCR \\
 & test\_date & date of COVID-19 test & dd/mm/yy \\
 & ctDate & date of CT-scan & dd/mm/yy \\
 & ctScore & CT value & number \\
 & ctScan & the participant had a CT-scan & True/False \\
 \bottomrule
\end{tabular}
}
\caption{Description of the meta-data collected and released as part of the Coswara dataset.}
\label{tab:collected_datasheet}
\end{table}
\subsection*{Sound recording specification}
\begin{figure}[!h]
    \centering
    \input{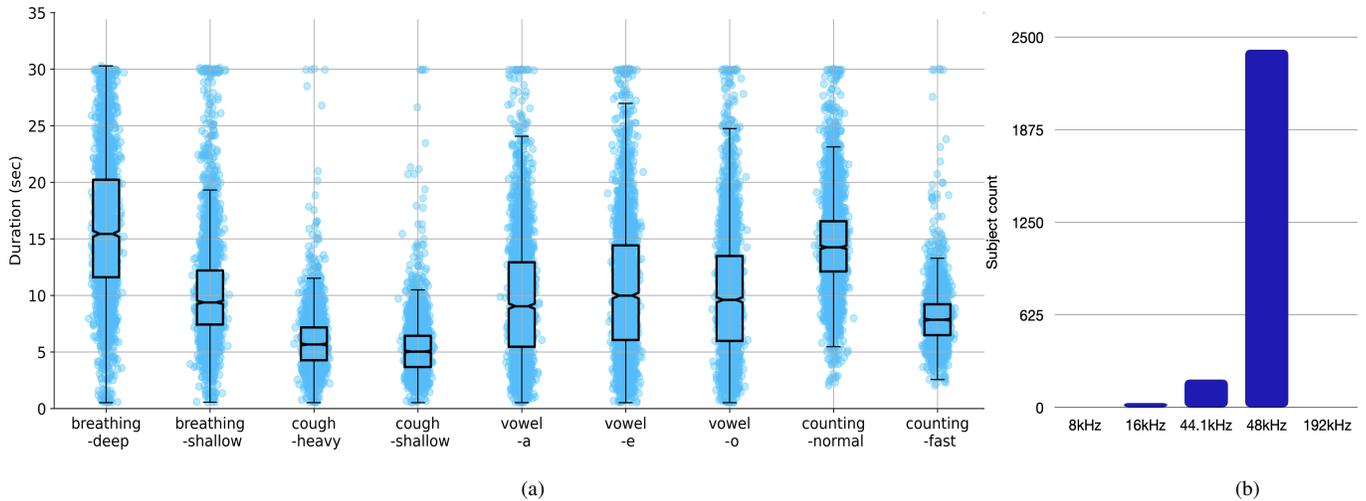}
    \vspace{-0.05in}
    \caption{Distribution of the (a) sound recording duration corresponding to the $9$ categories, (b) sampling rate of the devices used by the subjects.}
    \label{fig:plot_sound_recordings_stat}
\vspace{-0.1in}
\end{figure}

The sound files were recorded and stored in an  uncompressed audio format. Figure~\ref{fig:plot_sound_recordings_stat}(a) depicts the duration distribution of all the sound files after discarding the audio files of quality rating $3$ (poor quality,  as identified by the human listeners). The maximum duration in the recording setup was limited to $30$~s. 
Each dot in the figure corresponds to the duration of a single audio file. Additionally, a standard box plot representation is overlaid on the distribution to highlight  the median, $25^{th}$, and $75^{th}$ percentile etc. Figure 5 (a) indicates that, on average, the cough samples have smaller duration compared to breathing and sustained vowel phonation. Also, as expected, the counting-normal audio files are on average longer in duration than those corresponding to counting-fast.

Within this duration range, the nine sound categories had a different distribution. On average, the cough recordings had a relatively smaller duration (median of $5$s) while the breathing-deep recordings  had the longest duration (median of $15$s). The three vowel sounds corresponding to sustained phonation had a similar distribution. Further, a majority of the sound samples were recorded at $48$~kHz ($90\%$) (see Figure~\ref{fig:plot_sound_recordings_stat}(b))

\subsection*{Sound category classification}
The nine sound categories were chosen such that the excitation and physical state of the respiratory system is well captured. We validated the  complimentary nature of the acoustic sounds by building a nine-class classifier amongst the different sound samples. The classifier was trained on acoustic features extracted from the sound samples.  From each sound file, a $128$-dimensional averaged mel-spectrum vector was extracted. This mel-spectrum was computed by averaging the short-time mel-spectrogram obtained with a window of  $25$~msec duration, $10$~msec shift and with $128$ mel-spaced spectral filters \cite{librosa}.

\begin{figure}[t!]
    \centering
    \includegraphics[width=9.7cm, height=8.3cm]{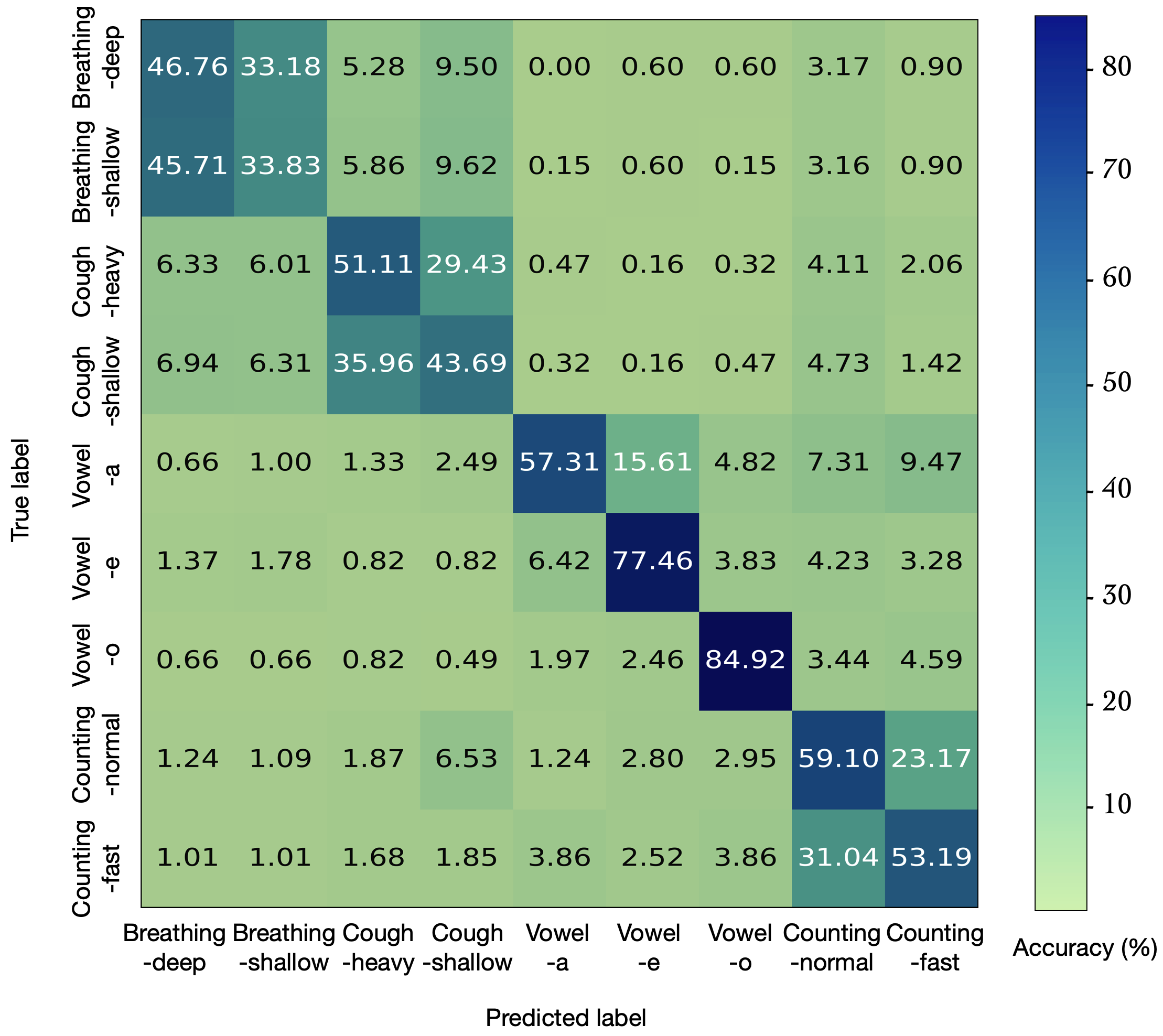}
    \caption{Confusion matrix for the audio category classification}
    \label{fig:confusion_matrix}
\end{figure}

A random forest classifier was trained using the acoustic feature data. A $70$-$15$-$15$\% train-val-test split made by (stratified) random sampling was done for training, validation and testing of the classifier, respectively.  The ``gini'' impurity criterion was used to train the classifier and the number of estimators was optimized using the validation dataset. A test set accuracy of $56.5\%$ was obtained. Fig.~\ref{fig:confusion_matrix} shows the resulting confusion matrix for the test data. Across all sound categories the accuracy was significantly greater than the chance performance of $11.1\%$. The confusion matrix also indicates a block diagonal like structure indicating that the class confusions mainly reside within the broad sound categories such as breathing, cough, and speech counting.

\subsection*{Gender classification from sound recordings}
\begin{figure}[t!]
    \centering
    \includegraphics[width=15.5cm, height=5.7cm]{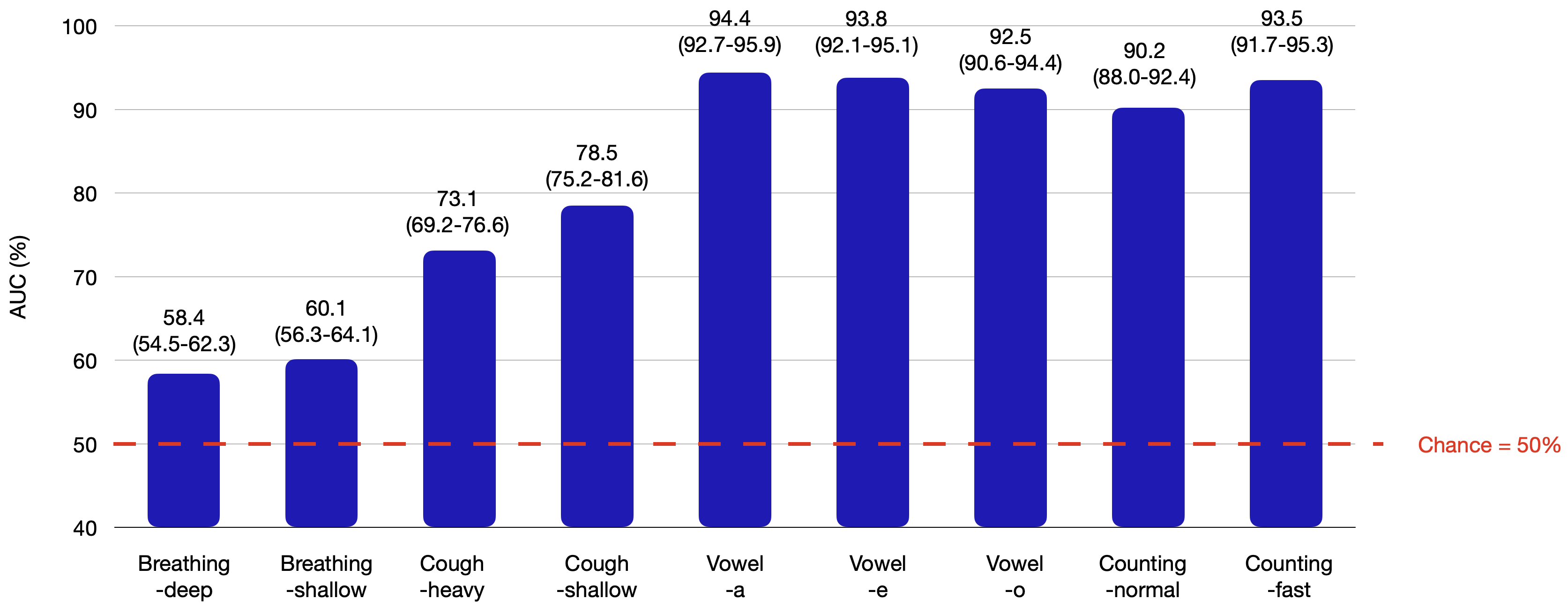}
    \caption{Test-set area under the curve (AUC) ($\%$) performance for the gender classification task from $9$ categories of respiratory sounds.}
    \label{fig:male-vs-female}
\end{figure}

According to speech production literature, gender prediction with speech sound recordings is a relatively easy task. In our dataset containing diverse sound samples, we validate the sound recordings for the presence of information related to the participant's gender. We performed a binary classification task, that is, male versus female, across all the sound categories. A stratified random sampling based $70-15-15$\% train-val-test split was used for training, validation and test dataset creation, respectively. A random forest classifier with ``gini'' impurity criterion was used. The area-under-receiver-operating-characteristic curve (AUC-ROC) was used as the performance measure. Figure~\ref{fig:male-vs-female} shows the classifier performance on the test set across different sound categories. The AUC was $\geq90\%$ for speech sound categories such as vowels and counting. Compared to speech, the performance degrades significantly for cough sound categories and for breathing sound categories.

\subsection*{COVID-19 detection from sound recordings}
Since the first release in April-2020, the Coswara dataset has gradually increased in size. The partitions  of the dataset have been used by different research groups \cite{9361107, coppock2021end, pal2021pay, alkhodari2022detection, pahar2022covid, ponomarchuk2021project, mohammed2021ensemble, sharma2021_dicova_1_csl, lella2022automatic, kranthi2022covid} for evaluating the possibility of COVID-19 detection using respiratory sound recordings. Further, two data challenges named, diagnosis of COVID-19 using acoustics (DiCOVA), were conducted using a subset of the data. These challenges invited participants to build machine learning systems for COVID-19 detection using the different sound categories \cite{sharma2021_dicova_1_csl, sharma2021_dicova_2_icassp}. These efforts demonstrated that the dataset can be used for developing machine learning algorithms for COVID-19 screening. 

In this paper, we trained 
the bi-directional long short term memory (BLSTM) based classifier  \cite{sharma2021_dicova_2_icassp} on the proposed Coswara dataset for the binary task of COVID/non-COVID classification. The classifier consisted of two BLSTM layers followed by a fully connected classification head. The classifier was trained on the segments extracted from the audio samples. The log-mel spectrogram was used as feature for training the classifier. The  classifier was trained using the weighted binary cross-entropy (BCE) loss.
Let, $N_{c}$ and $N_{nc}$ be the count of COVID and non-COVID subjects used in training, respectively. Let $r=N_{c}/N_{nc}$ be the class   ratio. Then,   the total loss is,
\begin{equation}
    L = -\sum _{i \in c} log(x_i) - r \sum _{i \in nc}   log(1-x_i) 
\end{equation}
where, $x_i$ denotes the probability of COVID-19 predicted by the model, and $c$ ($nc$)  denotes the set of COVID (non-COVID) samples.

\noindent The participants  having a moderate or excellent quality for the audio recordings, for all the nine categories, were selected. The final COVID probability score was obtained after linearly combining scores from the  classifiers trained on the nine different audio modalities and the symptoms. The filtered dataset was  randomly split into train, validation and test sets consisting of $238$, $58$ and $96$ COVID positive subjects, respectively and $707$, $178$ and $291$ non-COVID subjects, respectively. The area-under-the-receiver-operating-characteristics (AUC-ROC) was used as the evaluation metric as the dataset was imbalanced. 
The BLSTM classifier achieved an AUC of $91.5\%$[$95\%$ CI $88.5\%$-$94.1\%$] and sensitivity of $64.6\%$[$95\%$ CI $48.3\%$-$72.6\%$] calculated at $95\%$ specificity on the test set.
\begin{table}[t!]
\centering
\resizebox{\columnwidth}{!}
{
\begin{tabular}{llrcccclr}
\hline
                             &           & \multicolumn{1}{l}{}             & \multicolumn{1}{l}{} & \multicolumn{1}{l}{} & \multicolumn{1}{l}{} & \multicolumn{1}{l}{}                        &           & \multicolumn{1}{l}{} \\
\textbf{Subject sub-population}  & \textbf{} & \textbf{Count COVID / Non-COVID} & \textbf{}            & \textbf{AUC (\%)}    &                      & \textbf{Sensitivity at 95\% specificity} & \textbf{} & \textbf{p-value}     \\
                             &           & \multicolumn{1}{l}{}             & \multicolumn{1}{l}{} & \multicolumn{1}{l}{} & \multicolumn{1}{l}{} & \multicolumn{1}{l}{}                        &           & \multicolumn{1}{l}{} \\ \hline
                             &           & \multicolumn{1}{c}{}             &                      &                      &                      &                                             &           & \multicolumn{1}{l}{} \\
\textbf{Gender}              &           & \multicolumn{1}{c}{}             &                      &                      &                      &                                             &           & \multicolumn{1}{l}{} \\
Male                         &           & 51/209                           &                      & 92.7 (89.3-95.4)    &                      & 64.7 (37.3-76.7)                           &           & 0.901                \\
Female                       &           & 45/82                            &                      & 90.3 (85.1-95.1)     &                      & 64.4 (48.9-80.0)                            &           & 0.814                \\ \cline{1-1} \cline{3-3} \cline{5-5} \cline{7-7} \cline{9-9} 
                             &           &                                  &                      &                      &                      &                                             &           & \multicolumn{1}{l}{} \\
\textbf{Age}                 &           &                                  &                      &                      &                      &                                             &           &                      \\
15-30                        &           & 39/143                           &                      & 90.0 (85.1-94.3)     &                      & 59.0 (41.7-76.5)                           &           & 0.179                \\
30-45                        &           & 20/86                            &                      & 89.8 (80.8-96.4)     &                      & 55.0 (33.3-76.5)                            &           & 0.242                \\
45-60                        &           & 22/35                            &                      & 94.4 (89.6-98.6)     &                      & 68.2 (32.1-95.5)                          &           & 0.590                \\
60-90                        &           & 8/6                              &                      & 91.7 (75.0-100.0)    &                      & 50.0 (25.0-100.0)                           &           & 0.129                \\ \cline{1-1} \cline{3-3} \cline{5-5} \cline{7-7} \cline{9-9} 
                             &           &                                  &                      &                      &                      &                                             &           & \multicolumn{1}{l}{} \\
\textbf{Vaccination}         &           &                                  &                      &                      &                      &                                             &           & \multicolumn{1}{l}{} \\
With                         &           & 66/57                            &                      & 86.3 (80.8-91.3)     &                      & 31.8 (14.1-70.4)                            &           & p\textless{}0.050    \\
Without                      &           & 30/234                           &                      & 88.8 (82.3-93.9)     &                      & 53.3 (31.2-78.9)                           &           & 0.260                \\ \cline{1-1} \cline{3-3} \cline{5-5} \cline{7-7} \cline{9-9} 
                             &           & \multicolumn{1}{c}{}             &                      &                      &                      &                                             &           & \multicolumn{1}{l}{} \\
\textbf{Mask}                &           & \multicolumn{1}{c}{}             &                      &                      &                      &                                             &           & \multicolumn{1}{l}{} \\
With                         &           & 45/36                            &                      & 83.8 (76.1-90.6)     &                      & 46.7 (12.2-70.2)                            &           & p\textless{}0.050    \\
Without                      &           & 51/255                           &                      & 90.9 (86.4-94.9)     &                      & 64.7 (50.0-80.7)                           &           & 0.151                \\ \cline{1-1} \cline{3-3} \cline{5-5} \cline{7-7} \cline{9-9} 
                             &           &                                  &                      &                      &                      &                                             &           & \multicolumn{1}{l}{} \\
\textbf{English Proficiency} &           &                                  &                      &                      &                      &                                             &           & \multicolumn{1}{l}{} \\
With                         &           & 87/280                           &                      & 91.4 (88.4-94.2)     &                      & 58.6 (48.8-75.6)                            &           & 0.899                \\
Without                      &           & 9/11                             &                      & 86.9 (71.0-100.0)    &                      & 44.4 (16.7-100.0)                           &           & 0.379                \\ \cline{1-1} \cline{3-3} \cline{5-5} \cline{7-7} \cline{9-9} 
                             &           &                                  &                      &                      &                      &                                             &           & \multicolumn{1}{l}{} \\
\textbf{Date of Collection}  &           &                                  &                      &                      &                      &                                             &           & \multicolumn{1}{l}{} \\
Before Dec. 21               &           & 47/147                           &                      & 91.1 (86.6-94.8)     &                      & 55.3 (44.7-75.0)                            &           & 0.114                \\
After Dec. 21                &           & 49/144                           &                      & 92.3 (88.5-95.8)     &                      & 63.3 (38.6-78.2)                            &           & 0.123                \\ \cline{1-1} \cline{3-3} \cline{5-5} \cline{7-7} \cline{9-9} 
                             &           & \multicolumn{1}{c}{}             &                      &                      &                      &                                             &           & \multicolumn{1}{l}{} \\
\textbf{Collection Province} &           & \multicolumn{1}{c}{}             &                      &                      &                      &                                             &           & \multicolumn{1}{l}{} \\
Karnataka                    &           & 50/95                            &                      & 95.4 (91.6-98.3)     &                      & 82.0 (42.6-97.3)                            &           & 0.444                \\
Others                       &           & 46/196                           &                      & 88.5 (83.9-92.7)     &                      & 54.3 (38.6-74.4)                            &           & 0.568                \\ \cline{1-1} \cline{3-3} \cline{5-5} \cline{7-7} \cline{9-9} 
                             &           &                                  &                      &                      &                      &                                             &           & \multicolumn{1}{l}{} \\
\textbf{Prevalence}          &           &                                  &                      &                      &                      &                                             &           & \multicolumn{1}{l}{} \\
20\%                         &           & 72/288                           &                      & 90.9 (87.5-94.0)     &                      & 63.6 (45.6-72.7)                            &           & 0.814                \\
10\%                         &           & 32/288                           &                      & 91.2 (86.4-95.4)     &                      & 63.1 (40.3-77.0)                            &           & 0.783                \\
5\%                          &           & 15/285                           &                      & 88.9 (80.4-95.7)     &                      & 56.0 (29.1-75.2)                            &           & 0.574                \\
2.5\%                        &           & 7/273                            &                      & 88.8 (82.8-98.0)     &                      & 51.4 (34.0-87.6)                            &           & 0.542                \\ \hline
\end{tabular}
}
\caption{Description of the subject categories used in analysis, the AUC performance and the sensitivity at $95$\% specificity obtained using the BLSTM classifier. The p-values correspond to the t-test performed between scores from the full test set and the ones from the population sub-group considered.}
\label{tab:bias_analysis}
\vspace{-0.05in}
\end{table}
\subsection*{Analyzing bias in COVID-19 detection}
The diverse set of metadata, as described in Table~\ref{tab:collected_datasheet}, also allowed us to analyze the bias and fairness of the trained BLSTM based classifier. In order to analyze the dependence of the model performance on the metadata information, we divided the test set  into different population subgroups based on gender, age, vaccination status, mask usage, English proficiency, date of data collection, and province. 
The COVID-19 detection performance on these subgroups is reported in Table~\ref{tab:bias_analysis}. For all the AUC and sensitivity values reported, the two-sided $95\%$ confidence intervals (CIs) were calculated using bootstrap re-sampling with $1000$ bootstrap samples (sampled with replacement) \cite{carpenter2000bootstrap}. The p-values are obtained by using a t-test for comparing the scores obtained for the full test set and the scores for the population subgroups. 

\noindent The subject count varied across different subgroups. In the gender category, similar performance was observed for both male and female subgroups. The scores when compared with those obtained for the full test set were not significantly different.

\noindent The geographic location was split based on participants from within the Indian province of Karnataka and the rest of India. The resulting performance was not significantly different from the original full test set. In the original test-set, the COVID prevalence was 25\%. We experimented with different subsets of the test set by randomly selecting non-COVID samples to match a COVID prevalence level of 2.5\%, 5\%, 10\% and 20\%. As seen in Table~\ref{tab:bias_analysis}, the test sets with different prevalence levels did not generate statistically significant differences when compared with the original test set.  We analyzed the performance attained on two test sets created based on the time period of data collection, that is, chosen
 as a period before and after the onset of the Omicron variant in India. Note that the training data comprised of data samples
 collected before December, 2021. The model performance showed similar performance on samples collected after December,
 2021 (see Table~\ref{tab:bias_analysis}). This indicates that the timeline of data collection did not influence the  AUC performance.
 We created and analyzed population subgroups based on the vaccination status, and the use of mask during sound recording.
 The subgroup of subjects wearing a mask gave a performance that was significantly different compared to the full test set. This
 indicates that presence of a facial mask during the recording  modified the acoustic properties of the sound samples that were collected. Further, the subgroup
 of subjects who were vaccinated also had a significantly different performance. This can be attributed to the relatively milder
 symptoms observed in COVID subjects who were vaccinated.

\section*{Impact and Usability}
In this work, we have detailed the data collection and modeling efforts done as part of the development of a point-of-care testing (POCT) tool for screening of COVID-19. The data collection efforts spanned a period of 22 months and involved collaborative efforts from multiple hospitals, health centers, and the general public via crowd-sourcing. The rich set of acoustic stimuli, with 9 different categories of sounds consisting of breathing, cough, and speech, along with extensive metadata, allows the development of a COVID-19 screening tool. The designed website tool, using a BLSTM architecture,  was also made publicly available. The overall time spent by the participant to record the data ranged from $5$-$7$~mins. Thus, the proposed framework can provide rapid screening results without any additional sophisticated equipment. Further, the entire tool is developed for a smartphone based application, using models built on data collected from wide-range of cellphones. Hence, we hypothesize that the approach may generalize well for a massive population level use case. To the best of our knowledge, this study is a first of its kind to provide a publicly available web-tool for COVID-19 screening. Furthermore, we report a detailed bias and fairness analysis to understand the confounding factors associated with the COVID-19 screening tool developed using the collected data. As future extensions of this work, a collaborative effort can be taken to enlarge the subject count and also increase the diversity (and population size) associated with different demographics, gender, age groups, and health conditions.

Respiratory related diseases are a significant contributor to the leading causes of deaths \cite{salvi2018burden}.  In this context, automatic monitoring of the respiratory health can help recommend early interventions to an individual and hence, avoid deterioration in the respiratory health. Automatic monitoring by analyzing the respiratory sounds is particularly interesting owing to the ease in recording and analysis of the sound samples using mobile phones. It can also provide a scalable and cost-effective means for monitoring respiratory health. Providing a means to explore this direction, the Coswara dataset documents the respiratory sound samples drawn majorly from the Indian population. The subject population spans a wide age group (between 15-90 years) and the dataset contains an extensive meta-data for each subject. The sound samples contain two variants of breathing, two variants of coughs, three kinds of sustained vowel phonation, and two styles of continuous speech. A manual validation of the sound samples via human listening is also provided. The open access release of the dataset has drawn interest in the research community, and multiple studies have analyzed its use to explore COVID-19 detection \cite{9361107, coppock2021end, pal2021pay, alkhodari2022detection, pahar2022covid, ponomarchuk2021project, mohammed2021ensemble, sharma2021_dicova_1_csl, lella2022automatic, kranthi2022covid}. Complementing the few other COVID-19 respiratory sound sample datasets (comparison provided in Table 1), the Coswara dataset provides scope for development and detailed evaluation of AI-based screening tools. At a broader level, this will benefit design of explainable AI methodologies for respiratory health monitoring.

\section*{Code Availability}
\noindent The code used for technical validation is available at\\ \url{https://github.com/iiscleap/Coswara-Data/tree/master/technical_validation}.
\newpage
\bibliography{cas-refs}

\section*{Author Contributions}
\noindent SG, SRC, and NKS contributed in conceptualizing the study.  SG, SRC, NKS, and DB contributed to the study design. CC, SN, SKK, SG, MA, and PM contributed to data acquisition. DB, SRC, SG, NKS, and DD contributed to data analysis  and verified the data records. DB and NKS contributed to the initial drafting of the manuscript. All authors contributed to data interpretation and critical revision of the manuscript.  All authors had full access to all the data in the study and took responsibility for the decision to submit this draft for publication.

\section*{Competing Interests}
\noindent The authors declare no competing interests.

\section*{Acknowledgements}
\noindent We express our gratitude to Anand Mohan for the help in design of the web-based data collection platform. We also thank Prashant Krishnan, Ananya Muguli, Rohit Kumar and Dr Lancelot Pinto, for their contributions in the early part of the data collection.

\end{document}